\documentclass[10pt,twocolumn,letterpaper]{article}
\usepackage[accsupp]{axessibility}

\usepackage[pagenumbers]{cvpr} 

\usepackage{graphicx}
\usepackage{CJKutf8}
\usepackage{amsmath}
\usepackage{amssymb}
\usepackage{booktabs}

\usepackage[pagebackref,breaklinks,colorlinks]{hyperref}

\usepackage[capitalize]{cleveref}
\crefname{section}{Sec.}{Secs.}
\Crefname{section}{Section}{Sections}
\Crefname{table}{Table}{Tables}
\crefname{table}{Tab.}{Tabs.}

\def\cvprPaperID{*****} 
\def\confName{CVPR}
\def\confYear{2025}

\begin{document}
\begin{CJK}{UTF8}{gbsn}
\title{RepNet-VSR: Reparameterizable Architecture for High-Fidelity Video Super-Resolution}
\author{
Biao Wu\and
Diankai Zhang\and
Shaoli Liu\and
Si Gao\and
Chengjian Zheng\and
Ning Wang\and
State Key Laboratory of Mobile Network and Mobile Multimedia Technology, ZTE, China
\and
{\tt\small \{wu.biao,zhang.diankai,liu.shaoli,gao.si,zheng.chengjian,wangning\}@zte.com.cn}
}
\maketitle

\begin{abstract}
   As a fundamental challenge in visual computing, video super-resolution (VSR) focuses on reconstructing high-definition video sequences from their degraded low-resolution counterparts. While deep convolutional neural networks have demonstrated state-of-the-art performance in spatial-temporal super-resolution tasks, their computationally intensive nature poses significant deployment challenges for resource-constrained edge devices, particularly in real-time mobile video processing scenarios where power efficiency and latency constraints coexist. In this work, we propose a Reparameterizable Architecture for High Fidelity Video Super Resolution method, named RepNet-VSR, for real-time 4x video super-resolution. On the REDS validation set, the proposed model achieves 27.79 dB PSNR when processing 180p→720p frames in 103 ms per 10 frames on a MediaTek Dimensity NPU. The competition results demonstrate an excellent balance between restoration quality and deployment efficiency. The proposed method scores higher than the previous champion algorithm of MAI video super-resolution challenge. 
\end{abstract}


\section{Introduction}
\label{sec:intro}

Over recent years, convolutional neural networks (CNNs) have demonstrated unprecedented performance across diverse computer vision domains. Despite their empirical success, practical deployment of these models remains constrained by their substantial parameter counts and floating-point operations (FLOPs), typically requiring execution on cloud-based GPU servers. This limitation conflicts with growing industrial demands for on-device AI capabilities in mobile ecosystems, thereby driving cross-disciplinary research into efficient CNN deployment strategies for edge devices. When deploying AI-based solutions on mobile devices, it is important to pay attention to the characteristics of mobile NPUs and DSPs in order to design efficient models. \cite{Aibenchmark2019,Aibenchmark2018} provide an extensive overview of smartphone AI acceleration hardware and its performance. Based on the results reported in these papers, the latest mobile NPUs are close to the results of mid-range desktop GPUs published not long ago. Thanks to the development of edge chip computing power and edge AI accelerators, edge AI algorithms have flourished.

Video super-resolution (VSR) has emerged as a pivotal technology for enhancing visual content across industries, enabling the reconstruction of high-resolution (HR) video sequences from low-resolution (LR) inputs. Its applications span diverse domains: streaming platforms leverage VSR to upscale legacy content  while minimizing bandwidth consumption; surveillance systems utilize it to clarify critical details in low-light footage for forensic analysis; and mobile devices employ real-time VSR to refine zoomed-in videos or stabilize recordings without specialized hardware. In medical imaging, VSR enhances endoscopic video clarity for precise diagnostics, while autonomous vehicles rely on it to interpret distant objects in adverse weather conditions. Despite these transformative use cases, deploying VSR on resource-constrained edge devices, such as smartphones, drones, or IoT cameras, introduces significant technical hurdles.

In order to promote the development of video super-resolution on mobile terminals, MAI2025 has a real-time video super-resolution challenge track on an actual mobile accelerator. The challenge seeks to establish optimal equilibrium between reconstruction fidelity (quantified by PSNR) and operational efficiency (measured through end-to-end latency), requiring solutions to be able to balance accuracy-efficiency.

In this work, we propose a Reparameterizable Architecture for High Fidelity Video Super Resolution method, named RepNet-VSR, for real-time 4x video super-resolution. Demonstrating state-of-the-art efficiency-accuracy co-optimization, our reparameterized architecture achieves 27.79 dB PSNR on the REDS validation dataset for 4× video super-resolution (180p→720p) in 103 ms/f on a MediaTek Dimensity NPU. In summary,  our main contributions are as follows:

\begin{itemize}
    \item We use NAS search to search for the best parameter configuration between PSNR and FLOPs.
\end{itemize}
\begin{itemize}
    \item We employ 1×1 convolutional layers for dimensionality reduction, enhancing computational efficiency while preserving feature representation capabilities.
\end{itemize}

\begin{itemize}
    \item Multi-level feature fusion via channel concatenation enhances representational capacity by combining low-level textures and high-level semantics.
\end{itemize}

\begin{itemize}
    \item We enhance the representation capabilities of network through structural reparameterization, using more complex networks during training and merging them into a common convolution during inference to optimize representation capabilities and computational efficiency.
\end{itemize}

\section{Related Work}
\label{sec:RelatedWork}
In recent years, numerous efficient methods have been developed for image super-resolution tasks \cite{Fal,Fmen, FdI, Lisr}, demonstrating robust performance even on hardware with constrained computational capabilities. Following the pioneering work of SRCNN \cite{SRCNN}, which introduced convolutional neural networks (ConvNets) to super-resolution, FSRCNN \cite{FSRCNN} achieved substantial acceleration in single image super-resolution (SISR) networks by utilizing the original low-resolution input and reducing convolution kernel sizes. DRCN \cite{Dcnn} and DRRN \cite{DRRN} attempted to minimize model parameters through recurrent architectures but incurred heavy computational overhead. To address this limitation, IDN \cite{IDN} and IMDN \cite{IMDN} implemented streamlined information fusion strategies to lower both parameter counts and computational demands, whereas RFDN \cite{RFDN} introduced a residual feature distillation mechanism. The recent surge in transformer-based models has also led to the integration of self-attention (SA) \cite{ELAN} into lightweight SR frameworks. For instance, SwinIR \cite{SwIR} adopted window-based SA, while ESRT \cite{ESRT} devised an efficient transformer variant termed Efficient Multi-Head Attention (EMHA) for SISR. Despite their effectiveness in modeling long-range dependencies, these transformer-based approaches remain computationally intensive.

Video super-resolution (VSR) poses significant challenges due to the necessity of extracting and integrating complementary information from multiple misaligned video frames to achieve high-quality video restoration. A widely used strategy is the sliding-window framework \cite{RBNVSR, TDAN, EDVR}, which restores each video frame by leveraging neighboring frames within a limited temporal window. In contrast, recurrent frameworks aim to harness long-term dependencies through iterative propagation of latent features across sequential frames. Broadly speaking, these techniques \cite{EVSRTRLSP, BRCNMFSR, RSDN} achieve superior model compactness compared to sliding-window approaches. However, fundamental challenges persist in preserving long-range temporal dependencies and ensuring robust feature alignment across frames within recurrent architectures. BasicVSR\cite{BasicVSR} employs bidirectional propagation to holistically integrate temporal information across the full input sequence for enhanced reconstruction fidelity. Notably, it utilizes optical flow estimation to achieve precise spatiotemporal feature alignment through adaptive warping operations. In RRN \cite{RRN},  the model processes sequential frames through hidden state inputs, employing identity mapping in the state propagation to effectively retain texture information throughout the network architecture. Although the above methods achieve high reconstruction quality, they are inefficient on resource-constrained mobile devices.

Neural Architecture Search (NAS)\cite{NASFLNLN, FALSRNAS, MONAS} automates the design of deep learning architectures for image super-resolution, replacing manual trial-and-error approaches. Recent works demonstrate that NAS can discover lightweight, efficient, or high-performance SR networks tailored to specific constraints (e.g, computational cost, parameters, or PSNR/SSIM metrics). Recent advances in super-resolution have exploited structural reparameterization \cite{ding2021repvgg,liu2022ecbsr,chen2022repsr,zhang2023swinrep} to employ more complex networks during training and merge them into a single vanilla convolution at inference time.

\section{Approach}
\label{sec:Approach}
\subsection{Real-Time Video Super-Resolution Challenge Constraints}

The input tensor of model must process 10 subsequent video frames with a shape of [1×180×320×30], where:
\begin{itemize}
    \item The first dimension (1) represents the batch size.
\end{itemize}
\begin{itemize}
    \item The second and third dimensions (180×320) correspond to the height and width of the input frames from the REDS dataset.
\end{itemize}
\begin{itemize}
    \item The fourth dimension (30) combines the 3 RGB color channels and 10 frames (3 × 10 = 30 channels).
\end{itemize}

 The output tensor should have a shape of [1×720×1280×30], maintaining the same batch size and channel count while upscaling the spatial resolution to 720×1280.

 The final score for this challenge, as defined in Formula (1), is determined by two key metrics:
\begin{itemize}
    \item Reconstruction quality of the output results.
\end{itemize}
\begin{itemize}
    \item Model runtime performance on the target MediaTek NPU.
\end{itemize}

\begin{equation}
\begin{aligned}
Score(PSNR, runtime)=\frac{2^{2\cdot(PSNR-27)}}{runtime}
\end{aligned}
\end{equation}

The scoring formula demonstrates that doubling the computational efficiency on the target platform yields a quality improvement equivalent to a 0.5dB gain in PSNR. This necessitates prioritizing both image quality restoration and computational performance optimization in network design. To address this dual challenge, we implemented a systematic experimental approach. Our methodology begins with a comprehensive evaluation of state-of-the-art super-resolution techniques, followed by a detailed exposition of our novel algorithm's architectural advantages and performance enhancements.

\begin{figure*}[ht]
    \centering
    \includegraphics[width= 16cm]{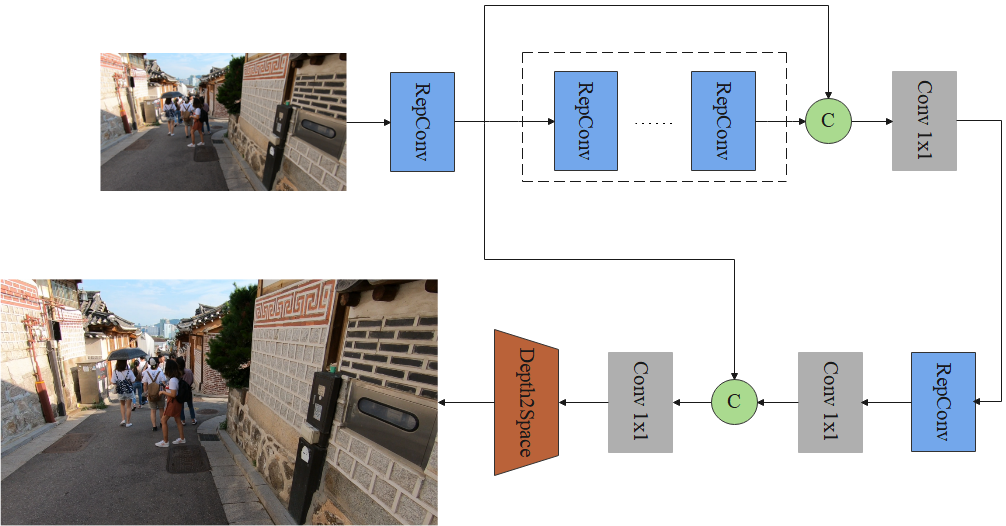}
    \caption{RepNet-VSR architecture overview.}
    \label{fig:my_diagram}
\end{figure*}
\begin{figure*}[ht]
    \centering
    \includegraphics[width= 10cm]{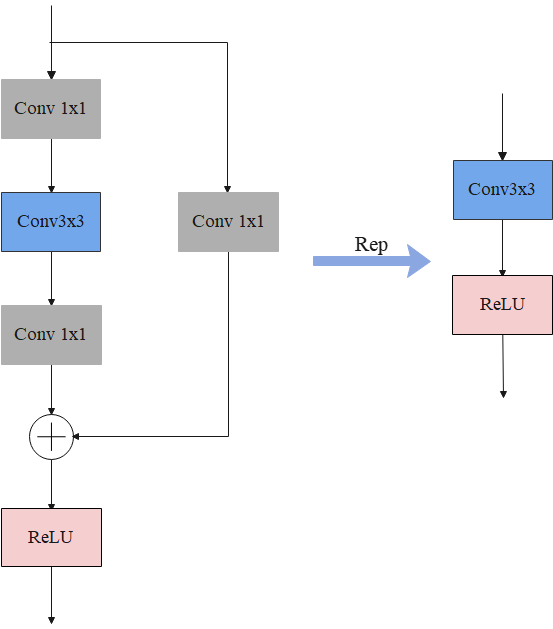}
    \caption{RepConv overview.}
    \label{fig:my_diagram}
\end{figure*}
\subsection{Analysis}
In previous iterations of the MAI Video Super-Resolution Challenge, participants demonstrated innovative approaches by developing computationally efficient algorithms that achieved an optimal equilibrium between visual reconstruction fidelity and runtime performance. The proposed EVSRNet\cite{EVSRNet} framework employs neural architecture search (NAS) technology to automatically optimize and identify optimal training parameters within its novel network architecture. Team Diggers \cite{MAI2021}introduced a bidirectional recurrent neural network (BRNN) architecture for video super-resolution, which leverages temporal information by incorporating feature maps from both preceding and subsequent frames to enhance the resolution reconstruction of the current target frame. The MVideoSR team\cite{MAI2022} developed an ultra-compact neural network architecture featuring merely four convolutional layers and a 6-channel configuration for efficient feature extraction. This lightweight model was optimized through a multi-stage training methodology, achieving unprecedented computational efficiency and the lowest recorded power consumption on designated hardware platforms. RCBSR\cite{MAI2022} adopts a re-parameterized ECB architecture that employs multi-branch network structures during the training phase, which are subsequently consolidated into a single 3×3 convolutional layer for inference. This design strategy effectively optimizes the balance between achieving higher PSNR metrics and maintaining computational efficiency in practical deployment. While these algorithms demonstrate notable performance advantages, they present specific operational limitations. The bidirectional recursive architecture achieves superior reconstruction quality, yet exhibits significant computational inefficiency during execution. Another prevalent issue arises in implementations employing channel compression through 3×3 convolution prior to depth-to-space operations. Particularly in 4× upscaling frameworks, this convolutional layer - designed to reduce feature channels from 48 to 3 - creates a critical processing bottleneck due to its position in the computational pipeline. The architectural requirement for high-dimensional feature transformation prior to spatial reorganization substantially impacts real-time processing capabilities.

\subsection{Network Architecture}
Motivated by these findings, we propose RepNet-VSR - a re-parameterizable architecture. This approach specifically addresses the identified computational bottlenecks while maintaining reconstruction fidelity through multi-level feature fusion.
As shown in Figure 1, we use channel concatenation to integrate multi-level features of deep and shallow network layers to enhance the feature representation capability of the model. To optimize computational efficiency, we incorporate 1×1 convolutional layers for dimensionality reduction throughout the fusion process. Specifically, prior to the depth-to-space operation, we substitute conventional 3×3 convolutions with 1×1 convolutions to achieve efficient channel compression from 48 to 3 dimensions. This architectural modification can speed up upsampling by a factor of 4 while improving reconstruction quality.

\subsection{Reparameterization}
Reparameterization is a technique in neural network design where models employ intricate architectures during training phase (e.g, multi-branch with parallel operations) that are subsequently transformed into equivalent but computationally efficient structures during deployment. This transformation usually involves combining multiple parameterized layers (e.g, convolutional blocks and skip connections) into a single convolutional layer. As shown in Figure 2, The proposed reparameterization module adopts a sequential architecture comprising three key operations. Initially, a 1×1 convolutional layer expands the channel dimension by a factor of 4, followed by a 3×3 convolutional layer that extracts spatial features in this elevated feature space. Subsequently, a 1×1 convolutional layer reduces the channel dimension back to its original size, completing the bottleneck structure. In order to maintain dimensionality consistency, a residual connection is implemented using 1×1 convolution. This architecture design effectively deepens the network while maintaining computational efficiency, thereby enhancing the feature extraction capabilities of the model. The synergistic combination of channel expansion/reduction and residual learning mechanisms ultimately strengthens the representational capabilities of the model without introducing significant computational overhead.
\subsection{Neural Architecture Search}
To achieve the best balance between reconstruction quality and runtime efficiency, we use FGNAS \cite{FGNAS} to automatically search between reparameterized modules and channels. The search space presented in Table 1 was systematically established, with a focused exploration of channel count variations and reparameterized module quantities to identify optimal parameter configurations within the proposed architectural framework. This method achieves an effective balance between computational efficiency and reconstruction fidelity through NAS.
\begin{table}
  \centering
  \resizebox{0.67\columnwidth}{!}{
  \begin{tabular}{c|cccc}
    \toprule
    Factor & Search Space \\
    \hline
Convolution types&Normal	\\
Convolution kernel sizes&3	\\
Activation functions&ReLU\\
The number of channels&0,1,..,32\\
The number of Repblock&0,1,..,8\\
    \bottomrule
  \end{tabular}
  }
  \caption{The search space of operations.}
  \label{tab:1}
\end{table}
\subsection{Loss}
FGNAS seeks to simultaneously maximize task performance accuracy and minimize computational resource consumption in the identified model. To achieve this dual objective, our formulation combines two key components: 1) a primary task-specific loss function that optimizes prediction quality, and 2) a regularization term designed to penalize computational overheads associated with network resources including parameter count, floating-point operations (FLOPs), and inference latency. In our work, FLOPs are incorporated as a regularizer to penalize network complexity, primarily because they offer a computationally efficient metric for optimization.The objective function is formally given by formula (2).
\begin{equation}
\begin{aligned}
\min_{\theta,\psi}L(\psi,\theta)+\lambda\cdot R(\psi)
\end{aligned}
\end{equation}
\section{Experiments}
\label{sec:Experiments}

In this part, we will describe the implementation details of our proposed method and report the results on the REDS validation set.

\subsection{Datasets}

For this challenge, we employ the REDS \cite{REDS} dataset, a widely used benchmark for traditional video super-resolution tasks due to its extensive content diversity and dynamic scene composition. Following established protocols, we partition the dataset into 240 videos for training, 30 for validation, and 30 for testing. Each video consists of 100-frame sequences, with every frame maintaining a native resolution of 1280×720 pixels at 24 frames per second. To create low-resolution inputs, the original videos undergo bicubic downsampling with a scaling factor of 4. The resultant low-resolution sequences serve as model input, while the corresponding original high-resolution frames function as the reconstruction target.
\begin{figure*}[ht]
    \centering
    \includegraphics[width= 16cm]{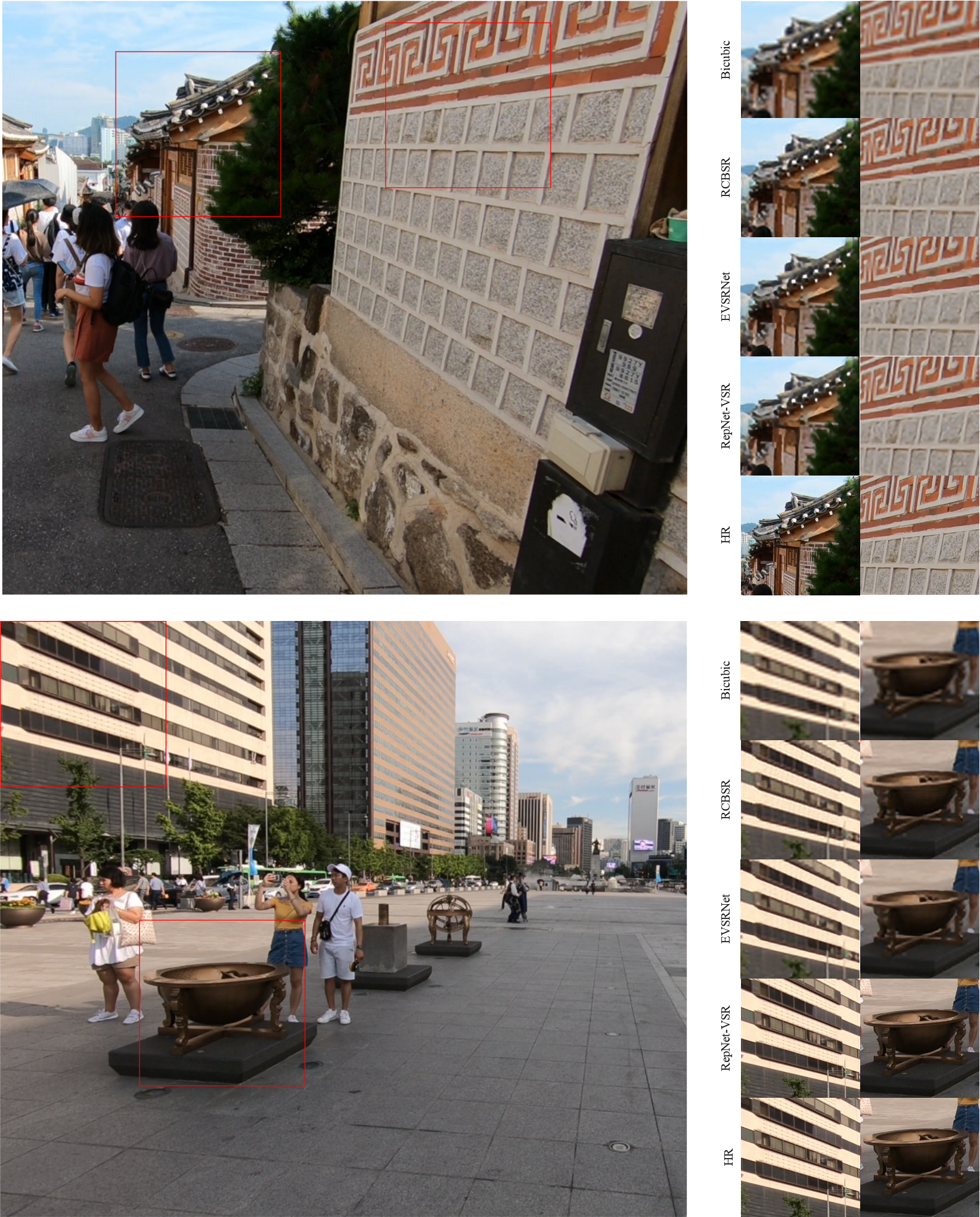}
    \caption{Qualitative comparison on the REDS val datasets. Zoom in for better visualization.}
    \label{fig:my_diagram}
\end{figure*}
\subsection{Training Configuration}
The models are trained using PyTorch Lightning with mixed precision (FP16) enabled by specifying the precision flag in the Trainer, along with the Adam optimizer configured with $\beta_1 = 0.9$ and $\beta_2 = 0.999$. We employ a initial learning rate of $5 \times 10^{-4}$. A decaying learning rate scheduler is implemented across all stages, featuring a 500-epoch warm-up period followed by linear decay until the learning rate reaches $1 \times 10^{-8}$. We randomly extract HR patches with dimensions 384×384 from high-resolution images and corresponding LR patches of size 96×96 from low-resolution images. The training process consists of two distinct phases. During the initial phase, Neural Architecture Search (NAS) is employed to conduct architectural search, optimizing both channel quantities and the count of re-parameterizable modules. The objective function defined in Equation 2 guides this optimization to maintain an optimal equilibrium between image restoration quality (PSNR) and computational efficiency. In the subsequent optimization phase, the best configuration determined in the first phase is retained, and the reconstruction quality of the pre-trained model is adjusted using L2 Loss fine-tuning. All hyperparameters remain consistent in both phases to ensure the continuity of training, and the second phase focuses on further performance optimization on the established architectural framework.

\begin{table}
  \centering
  \resizebox{1.0\columnwidth}{!}{
  \begin{tabular}{cccccc}
    \toprule
    Model & nc & nb & PSNR & Runtime(ms) & Score \\
    \midrule
    {1} & 8 & 4 & 27.39 & 56.5 & {0.0303} \\
    \textcolor{blue}2 & \textcolor{blue}{16} & \textcolor{blue}4 & \textcolor{blue}{27.79} & \textcolor{blue}{89.6} & \textcolor{blue}{0.0334} \\
    3 & 32 & 4 & 28.01 & 149.7 & 0.0271 \\
    4 & 16 & 5 & 27.83 & 93.6 & 0.0323 \\
    \bottomrule
  \end{tabular}
  }
  \caption{Training results of different models on the REDS validation set.}
  \label{tab:example}
\end{table}

\subsection{Runtime evaluation}
As previously noted, while our training pipeline utilizes PyTorch Lightning, the efficiency evaluation necessitates a TensorFlow Lite (tflite) model. To achieve this, we transfer the PyTorch model weights into an architecturally equivalent TensorFlow implementation by performing dimension transformations across network layers, ultimately generating a floating-point precision tflite model for runtime evaluation. To check the validity of the model, the challenge requires running it using AI Benchmark on a MediaTek-based device, using FP16 mode + MediaTek neuron delegate; if there are other platforms, use NNAPI / Qualcomm QNN delegate), as the final runtime evaluation will use a similar setup. Given the current unavailability of the specified hardware infrastructure, we are conducting computational efficiency assessments of the selected model using FP16 precision optimization on the Qualcomm Snapdragon 870's GPU platform. This interim evaluation will be followed by formal coordination with the MAI2025 organizing committee to perform standardized efficiency benchmarking through their designated testing protocols.

\subsection{Ablation Studies}
We employ a two-phase training strategy. During the initial phase, Neural Architecture Search (NAS) is utilized to conduct preliminary parameter optimization regarding channel quantities and the number of reparameterization modules. Subsequently, we initialize the model with parameters from the preceding phase and perform fine-tuning using an L2-norm loss function for enhanced performance refinement. As indicated in Table 2, we evaluated the computational efficiency of processing 10 frames using FP16 precision mode on the Qualcomm Snapdragon 870 GPU platform. In addition, Table 2 shows several sets of parameter configurations with relatively high scores under the proposed network architecture. Experimental results indicate that the optimal configuration under the proposed architecture occurs with 16 channels and 4 reparameterization modules.

\subsection{Test on MediaTek's NPU}
In the final phase of our evaluation, we conducted a comprehensive efficiency assessment on the dedicated Neural Processing Unit (NPU) of MediaTek.

\begin{itemize}
    \item Quality-PSNR Progression:
    
    Our model (MAI2025) achieves a 1.8\% improvement in PSNR over EVSRNet (27.79 vs. 27.42) and a 1.9\% improvement over RCBSR [34] (27.79 vs. 27.28), setting a new standard for reconstruction quality and outperforming RCBSR based on ECB reparameterization.
\end{itemize}

\begin{itemize}
    \item Computational Efficiency:
    CPU Utilization:
    
    Our solution shows comparable CPU latency to EVSRNet (273ms vs 271ms), though 2.44× slower than RCBSR's CPU implementation (273ms vs 112ms) 
    NPU Acceleration: 
    
    Maintains parity with EVSRNet (103ms) while achieving 7.5\% slower inference than RCBSR (103ms vs 95.8ms) on the NPU of MediaTek.
\end{itemize}
\begin{itemize}
    \item Score Metric Analysis:
    
    Our model achieves a 66.7\% higher composite score than EVSRNet (0.029 vs. 0.0174) and 88.3\% improvement over RCBSR.
\end{itemize}

From Table 3, it can still achieve a better quality-efficiency trade-off despite longer CPU runtime.
\begin{table}
  \centering
  \resizebox{1.0\columnwidth}{!}{
  \begin{tabular}{cccccc}
    \toprule
    Model & year & PSNR & CPU Runtime & NPU Runtime & Score \\
    \midrule
    {EVSRNet} & MAI2021 & 27.42 & 271 & 103 & {0.0174} \\
    RCBSR & MAI2022 & 27.28 & 112 & 95.8 & 0.0154 \\
    \textcolor{blue}{our} & \textcolor{blue}{MAI2025} & \textcolor{blue}{27.79} & \textcolor{blue}{273} & \textcolor{blue}{103} & \textcolor{blue}{0.029} \\
    \bottomrule
  \end{tabular}
  }
  \caption{Inference time (ms) of RepNet-VSR on MediaTek NPU.}
  \label{tab:example}
\end{table}
\subsection{Visual Comparision}
Our proposed RepNetVSR algorithm was evaluated against bicubic interpolation, EVSRNet, and RCBSR in terms of subjective visual quality, with comparative results demonstrating its significant superiority in perceptual performance, as shown in Figure 3.

\section{Conclusion}
\label{sec:Conclusion}

In this paper, we introduce RepNet-VSR, a reparameterizable architecture designed to optimize inference efficiency while maintaining high restoration quality in video super-resolution (VSR). Our method leverages reparameterization to streamline computational overhead during deployment, enabling real-time performance on mobile devices. When evaluated on the REDS validation set, RepNet-VSR achieves a PSNR of 27.79 dB for 4x super-resolution (180p→720p), processing each 10 frames in 103 ms on a MediaTek Dimensity NPU. These results underscore its exceptional balance between reconstruction fidelity and computational efficiency, addressing the critical challenge of latency-sensitive mobile device applications. In addition, RepNet-VSR surpasses previous competitive algorithms in the MAI Video Super-Resolution Challenge, demonstrating its practical advantages in deployment scenarios.

\end{CJK}
\end{document}